\documentclass[aps,prb,showpacs,twocolumn,groupedaddress]{revtex4-1}
\usepackage{graphicx}
\usepackage{comment}
\begin{document}
\def\vector#1{\mbox{\boldmath $#1$}}
\title{Study of Localized Character of 4$f$ Electrons and Ultrasonic Dispersions in SmOs$_4$Sb$_{12}$ \\
by High-Pressure High-Frequency Ultrasonic Measurements}
\author{S. Mombetsu}
\author{T. Murazumi}
\author{H. Hidaka}
\author{T. Yanagisawa}\thanks{Corresponding author: tatsuya@phys.sci.hokudai.ac.jp}
\author{H. Amitsuka}
\affiliation
{Department of Physics, Hokkaido University, Sapporo 060-0810, Japan}
\author{P.-C. Ho}
\affiliation
{Department of Physics, California State University, Fresno, CA 93740, U.S.A.}
\author{M. B. Maple}
\affiliation
{Department of Physics and Center for Advanced Nanoscience, University of California San Diego, La Jolla, CA 92093, U.S.A.}
\date{\today}
\begin{abstract}
We present high-frequency ultrasonic measurements on the filled skutterudite SmOs$_4$Sb$_{12}$ under hydrostatic pressure. The results clarify that the 4$f$ electrons in this compound transform from delocalized at ambient pressure to localized at high pressures with a crossover pressure of approximately 0.7 GPa. 
This drastic change in the 4$f$ electrons under pressure is apparently related to the non-Fermi liquid state, which appears in an intermediate-pressure range of 0.5-1.5 GPa.
The results or our analysis strongly suggest that the ferro-octupolar interaction becomes dominant at high pressure.
Moreover, we report the pressure dependence of the ultrasonic dispersion, which is due to rattling, over a wide range of ultrasonic frequencies up to 323 MHz.
The drastic change in the ultrasonic dispersions and the frequency dependent elastic anomaly in the $C_{11}$ mode at lower temperatures imply a possible coupling between rattling phonons and 4$f$ electrons.
\end{abstract}
\pacs{71.27.+a, 62.65.+k, 62.20.de, 07.35.+k}
\maketitle
A large variety of quantum phenomena have been found in rare earth based intermetallic compounds, such as heavy-fermion (HF) behavior, multipolar order, and unusual types of superconductivity \cite{CeCu2Si2,CeCu6_Shiina,Recent_HF}. The $c$-$f$ hybridization ($i$.$e$., hybridization between conduction electrons and 4$f$-electrons in rare-earth ions) plays an important role in these systems. The $c$-$f$ hybridization causes two competitive effects: the Kondo effect and the Ruderman-Kittel-Kasuya-Yosida (RKKY) interaction. The Kondo effect leads to itinerant 4$f$ electrons that lose their local moments at low temperature. In contrast, the RKKY interaction, which is an intersite interaction between 4$f$ electrons via conduction electrons, causes an ordering of local multipole moments. Especially for systems based on Ce(4$f^1$) and Yb(4$f^{13}$), which have one electron or one hole in the 4$f$ orbital, these two effects compete with one another and are well described by the Doniach phase diagram, which is a good guide for understanding this competitive feature \cite{Doniach}. The strength of $c$-$f$ hybridization can be controlled by external parameters, such as magnetic field, chemical substitution, or pressure. For Ce compounds, when a magnetic order temperature is suppressed towards absolute zero by an external parameter, unusual physical properties, such as non-Fermi liquid (NFL) behavior and superconductivity \cite{CeCu2Ge2,CeCu2,CeRhIn5}, have been found in the vicinity of the quantum critical point. However, the $c$--$f$ hybridization and its quantum criticality for the {\it "plural"} 4$f$-electron systems ($i$.$e$., systems with more than two electrons in the 4$f$ orbital such as Pr(4$f^2$), Nd(4$f^3$) and Sm(4$f^5$ or 4$f^6$)) remain poorly understood.

\begin{figure*}[t]
\includegraphics[width=0.9\linewidth]{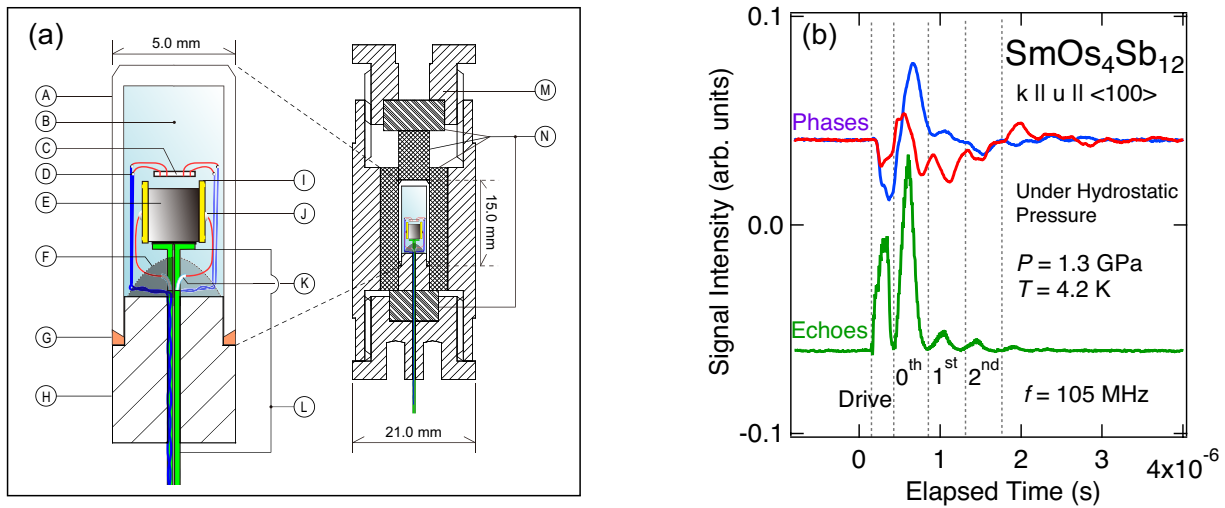}%
\caption{\label{fig:epsart} (a)  Schematic illustrations of hydrostatic-pressure ultrasonic measurement setup. A: Teflon capsule, B: Daphne oil 7373 for pressure medium, C: tin stick for a manometer, D: Cu and Au line, E: sample, F: Stycast 2850, G: Cu-Be sealing ring, H: body of a CuBe plug, I: LiNbO$_3$ transducer, J: Ag paste, K: Teflon insulation shield and inner conductor of semi-rigid coaxial cable, L: Cu outer shield of semi-rigid coaxial cable, M: Cu-Be parts (nuts and cylinder), N:  NiCrAl cylinder and WC pusher. (b) Ultrasonic echo signal (green) and phase signals (blue and red) of longitudinal $C_{11}$ mode at 105 MHz at 1.3 GPa, 4.2 K.}
\end{figure*}

Recently, rare-earth-based compounds with a highly symmetrical cage-like crystal structure have been attracting attention because they are prone to develop $c$-$f$ hybridization and/or low-lying highly degenerate crystalline-electric-field (CEF) states, which potentially have multipolar degrees of freedom. Such states have been observed in plural 4$f$-electron systems. In Pr-based cage-structured compounds, several novel physical properties have been observed, such as the quadrupolar Kondo effect in PrV$_2$Al$_{20}$(which is yet to be confirmed)[\onlinecite{PrV2Al20}], HF behavior, the unusual superconductivity of PrOs$_4$Sb$_{12}$ [\onlinecite{PrOs4Sb12}], and the unusual charge order in PrRu$_4$P$_{12}$ [\onlinecite{PrRu4P12}]. Novel physical properties are also observed in Sm-based compounds, for example, a magnetically ordered state that is unusually insensitive to external magnetic field and HF behavior in Sm$T_2$Al$_{20}$ ($T$ = Ti, V, Cr) \cite{SmTi2Al20, SmTr2Al20}. 
The origin of the magnetically insensitive state and the role of $c$-$f$ hybridization in such Sm-based compounds have not yet been clarified. 

The filled skutterudite SmOs$_4$Sb$_{12}$ with cubic space group  Im$\bar3$ (T$_h^5$, No. 204) is a good candidate for studying the $c$--$f$ hybridization effect of a Sm-based system. This compound has a large Sommerfeld coefficient $\gamma$ of approximately 820 mJ mol$^{-1}$ K$^{-2}$, which indicates HF behavior at low temperature \cite {Yuhasz,Sanada}. This value for $\gamma$ remains almost constant up to 8 T [\onlinecite{Sanada}], in contrast to typical Ce-based HF compounds \cite{CeCu6}, where the Kondo effect is weakened when subjected to an external magnetic field. The characteristic temperature $T^{\ast}$, below which 4$f$ electrons become itinerant, is estimated to be approximately 20 K for SmOs$_4$Sb$_{12}$ in zero magnetic field \cite{NQR_PRL}. Below $T^{\ast}$, this material undergoes a phase transition at $T_{\rm C} \approx 2.6$ K, accompanied by a weak spontaneous ferromagnetic (FM) moment of 0.02-0.03 ${\mu}_{\rm B}$ per Sm ion \cite {Yuhasz,Sanada}. Because this value is much smaller than that of the magnetic moment of a free Sm$^{3+}$ ion (0.71 ${\mu}_{\rm B}$), the ordered state is considered to be due to itinerant FM order. 
The average valence of the Sm ion is approximately +2.83 at 150 K, which decreases with decreasing temperature and levels off to a constant value of approximately +2.76 near $T^{\ast}$ [\onlinecite{Valence}].
Such anomalous coexisting of the mid-valence state and heavy Fermion state in SmOs$_4$Sb$_{12}$ could possibly be due to the coincidence of weak $c$-$f$ hybridization and the small energy difference between Sm divalent and trivalent states. \cite{Yamasaki} On the other hand, the log $T$ behavior on the temperature dependence of the valence change indicates an unconventional Kondo effect associated with Sm 4f-electron charge degrees of freedom. \cite{Fushiya}
A large anharmonic vibration of a guest ion, known as rattling, has been detected in this material as well as in PrOs$_4$Sb$_{12}$ and other skutterudites. Reference \onlinecite{Rattling_HF} discusses the possible contribution to the HF state of charge fluctuations due to rattling of the rare-earth ion.

Upon applying hydrostatic pressure to SmOs$_4$Sb$_{12}$, $T_{\bf C}$ increases whereas $T^{\ast}$ decreases \cite{NQR_PRL}. These opposite tendencies imply a recovery of the localized character of 4$f$ electrons when under pressure.
The jump in the specific heat ${\Delta}C/T_{\bf C}$ which occurs at $T_{\bf C}$, becomes larger with increasing pressure, which suggests that the ordered state is stabilized under pressure. In contrast, the maximum in the ac susceptibility at $T_{\bf C}$ becomes smaller under pressure and seems to disappear at higher pressures \cite{UnderP_C_AcChi}. These results imply that the ordered state at high pressures is not a simple FM ordered state but a higher-rank multipolar type of order. 
Thus, the possible multipolar contribution in SmOs$_4$Sb$_{12}$ should be investigated by using the appropriate probe for detecting the higher-rank multipoles.

By using ultrasonic measurements, the response of electric-quadrupole moments of localized 4$f$ electrons can be obtained spectroscopically \cite{QS}. In fact, ultrasonic measurements have detected the possible effects of multipoles in numerous rare-earth based compounds \cite{SmRu4P12,AFQ_PrIr2Zn20}. Recently, ultrasonic measurements of SmOs$_4$Sb$_{12}$ in a pulsed magnetic field have shown clear evidence of the ${\Gamma}_{67}$-quartet CEF ground state \cite{SmOs4Sb12_Pulse}. 
These results also suggest the existence of a Kondo-like screened state and its suppression by strong magnetic fields. Furthermore, ultrasonic dispersions, $i$.$e$., frequency-dependent elastic anomalies accompanied by a peak in ultrasonic attenuation, appear near 20 and 55 K in SmOs$_4$Sb$_{12}$ [\onlinecite{SmOs4Sb12_Yana}]. 
These anomalies are considered to be resonances between thermally activated rattling and ultrasonic strain. Thus, ultrasonic measurements with $c$-$f$ hybridization controlled by pressure should provide new information on multipoles, rattling and their consequences on the exotic HF state of SmOs$_4$Sb$_{12}$. Here we present the results of high-frequency ultrasonic measurements on SmOs$_4$Sb$_{12}$ under hydrostatic pressure up to 1.3 GPa.

In this study, we used single-crystalline SmOs$_4$Sb$_{12}$ grown using the Sb-self flux method. The single-phase nature of the present sample was confirmed by X-ray diffraction. Crystalline quality of the specimen was also checked by electrical resistivity and magnetic susceptibility at ambient pressure, which are consistent with previous reports. A well-isolated ultrasonic echo signal with frequency of 105 MHz (wavelength of $\approx40\mu$m) in the present crystal as seen in Fig. 1(b) is also a good benchmark to evidence the bulk sample's homogeneity. To generate and detect ultrasonic waves, LiNiO$_3$ transducers were fixed on the sample by using superglue (GEL-10: Toagosei Co. Ltd.).
To carry high-frequency signals under hydrostatic pressure, two ultrathin semi-rigid coaxial cables (SC-033/50: Coax Co. Ltd.), which comprise Cu-shield sleeves with a diameter of 0.33 mm$\phi$ and Au-coated W-core cables with impedance-adjusted (50 $\Omega$) diameters of 0.08 mm$\phi$, were installed into the pressure cell. 
As shown in Fig. 1(a), the sample was mounted on the delaminated outer Cu conductors of the coaxial cables, which served as thermal anchors between the sample in the pressure cell and the thermometers outside the pressure cell. 
The edge of the dielectric (PTFE) and shield tubes (Cu) were carefully sealed with Stycast 2850 to prevent pressure from leaking.
With this setup, we obtained clear ultrasonic-echo signals under a high pressure of 1.3 GPa at 4.2 K [see Fig. 1 (b)]. Thus, this setup allowed us to make ultrasonic measurements at variable frequencies up to 323 MHz even under high pressure.

\begin{figure}[t]
\includegraphics[width=0.9\linewidth]{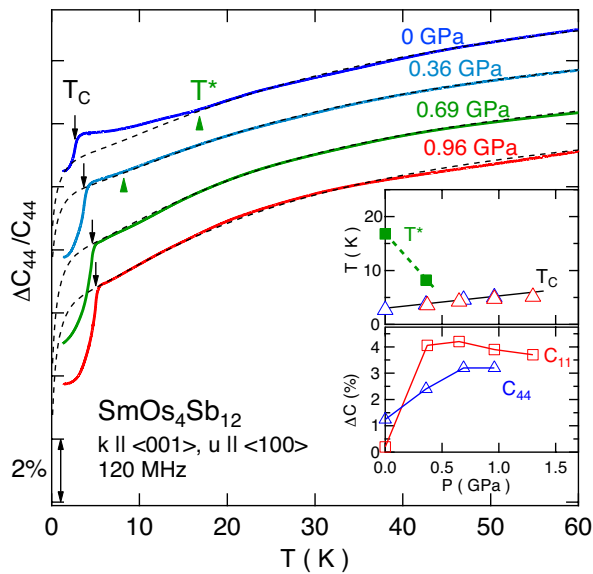}
\caption{\label{fig:epsart} Normalized elastic constant $C_{44}$ versus temperature measured at various hydrostatic pressures (solid lines). The data are shifted vertically for clarity. Black dashed curves represent the calculated elastic constant. The upward and downward arrows indicate the characteristic temperature $T^{\ast}$ and the ordering temperature $T_{\rm C}$, respectively. The upper inset shows a $T$--$P$ phase diagram and the lower inset shows the jump in the elastic constants at $T_{C}$ as a function of hydrostatic pressure.
}
\end{figure}

Figure 2 shows the relative change in the elastic constant $C_{44}$ as a function of temperature measured at various pressures.  Below 80 K, $C_{44}$ decreases with decreasing temperature ($i$.$e$., softening) at all pressures studied. This softening can be explained by the strain-quadrupole coupling. The black dashed curves in Fig. 2 represent the calculated relative change $C_{44}$ = $C_{44}^0-\frac{Ng^2{\chi}_{{\Gamma}_4}}{(1+g^{\prime}{\chi}_{{\Gamma}_4})}$, where $C_{44}^0$, $N$, $g$, ${\chi}_{{\Gamma}_4}$ and $g^{\prime}$ are the background elastic constant, the number of magnetic ions per unit volume, the quadrupole-strain coupling constant, the single-ion quadrupolar susceptibility and the intersite quadrupole-quadrupole coupling constant, respectively \cite{QS}. In the present analysis, we assume that the mixed valence state makes little contribution on the calculation, since a small pressure dependence of the averaged valence at lowest temperature within 0.04\% up to 0.9 GPa is reported \cite{Valence_Pdep}. Here we discuss only the relative change in the elastic constants as a function of temperature and do not determine the absolute value of $C_{44}^0$ and $g$. 
The intersite quadrupole-quadrupole coupling constant $g^{\prime}$, is negative at all of the pressures, which suggests antiferro quadrupolar interaction. $g^{\prime}$ is $-$1.55 and $-$1.50 K at ambient pressure and at 0.35 GPa, respectively, and at 0.69 and 0.96 GPa, $g^{\prime}$ is $-$0.90 K. 
The value of $C_{44}$ calculated on the basis of the localized 4$f$-electron model with the CEF ground state ${\Gamma}_{67}$(0 K)--${\Gamma}_{5}$(20 K) reproduces well the softening above the characteristic temperature $T^{\ast} \approx$ 17 K.
However, below  $T^{\ast}$, the calculated curve deviates from the measured data. This deviation can be explained by the suppression of the quadrupole moments of the 4$f$ electrons due to the $c$--$f$ hybridization that strengthens with decreasing temperature. Thus, we can treat $T^{\ast}$ as the crossover temperature between the delocalized and localized electron state. Under a hydrostatic pressure of 0.36 GPa, the calculation can reproduce the softening of $C_{44}$ down to 8 K. Such a decrease in $T^{\ast}$ as determined in this study is roughly consistent with the electrical resistance and nuclear-magnetic-resonance measurements under hydrostatic pressure, which leads to an estimate of $T^{\ast}$ that decreases slowly with increasing pressure \cite{NQR_PRL}. Above 0.7 GPa, the calculation appropriately reproduces all the features of $C_{44}$ down to $T_{\rm C}$. Thus, we conclude that the 4$f$ electrons recover their localized character at 0.7 GPa.

Such well-defined localized character of Sm under pressure advocated by present data is, however, inconsistent with the valence determined with X-ray absorption spectroscopy (XAS). The pressure dependence of the averaged valence determined with XAS methods shows no significant change at low temperature up to 3.5 GPa \cite{Fushiya}, in spite of a more significant effect of pressure on the valence would be expected if an oversized cage causes the mid-valence state of Sm in the present compound. On the other hand, a relatively small change of the 'temperature dependence' of the valence has been seen at around 1 GPa, at which the rattling of Sm ion also seems to be strongly affected (as discussed later in the present paper). Thus, it is obvious that ultrasonic measurement is rather sensitive for the detection of the 4f-electron's local charge degrees of freedom. At this moment, it is still controversial whether such apparent contradiction comes from an unconventional Kondo effect due to rattling \cite{Rattling_HF} or misinterpretation of either experimental results.

\begin{figure*}[t]
\includegraphics[width=0.8\linewidth]{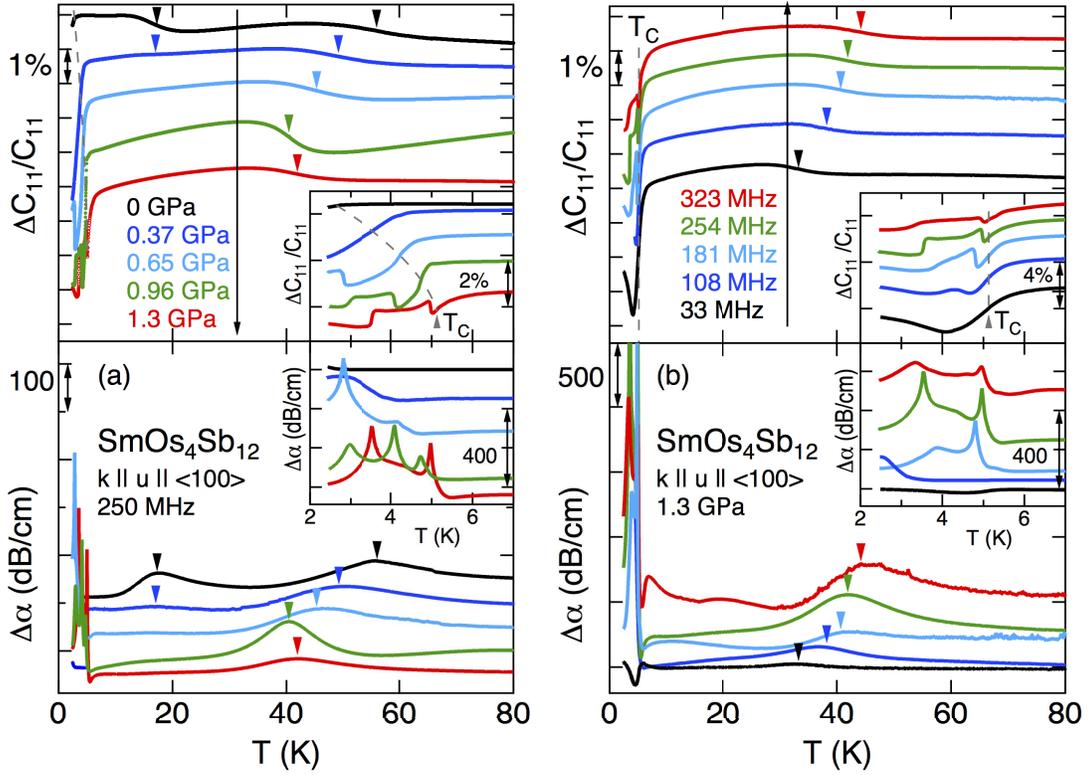}
\caption{\label{fig:epsart} Normalized elastic constant of $C_{11}$ vs temperature (a) measured at 250 MHz and under various hydrostatic pressures and (b) measured at 1.3 GPa and at various frequencies. The downward arrowheads indicate the temperatures at which the resonance relation, ${\omega}{\tau} \sim 1$ is satisfied. The insets give enlarged views of the low-temperature range. The lower panels show the corresponding ultrasonic attenuations. The data have been shifted vertically for clarity.}
\end{figure*}

Controversy of the valence aside, the present results of $C_{44}$ under pressure still provides further confirmation that the CEF ground state of the present compound is the ${\Gamma}_{67}$ quartet because the $\Gamma_5$ CEF-ground-state model predicts that the softening of the $C_{44}$ mode levels off below $\approx$ 10 K \cite{SmOs4Sb12_Yana}. In the present analyses, however, we assume that the CEF level splitting ($\Delta \approx 20$ K at ambient pressure) remain unchanged in the hydrostatic pressure below 1 GPa for simplification. If the level splitting $\Delta$ is set as a free parameter, alternatively the inter-site quadrupole-quadrupole coupling constant needs to be changed drastically in order to reproduce the temperature and pressure dependence of $C_{44}$ regarding the deviation at $T^*$. Therefore, further measurements are needed to get more information of the pressure dependence of the CEF splitting and $c$-$f$ hybridization.

Figure 3(a) shows the relative change in the elastic constant $C_{11}$ as a function of temperature measured at 250 MHz and under various hydrostatic pressures. At ambient pressure, $C_{11}$ reveals shoulder-like anomalies near 15 and 55 K, which are ultrasonic dispersions (UDs), and an abrupt decrease near $T_{\rm C}$ due to magnetic ordering, as described in Ref. \onlinecite{SmOs4Sb12_Yana}. The elastic anomalies of the UDs reveal a clear dependence on pressure. Between these UDs, $C_{11}$ softens, which is also due to the strain-quadrupole coupling. This softening must become less stronger below the crossover temperature $T^{\ast} \approx$ 17 K in the same manner as does $C_{44}$, although the presence of the UD near 15 K makes it difficult to distinguish the crossover phenomena in the $C_{11}$ mode.
Upon applying pressure, the softening of $C_{11}$ becomes larger and continuous down to $T_{\rm C}$, which is again explained by a recovery of the localized character of the 4$f$ electrons, as discussed for $C_{44}$.
The relatively small upturn near 15 K at ambient pressure becomes smaller upon applying pressure and seems to disappear near 0.65 GPa (see inset of Fig. 3(a)).
Below $T_{\rm C}$, $C_{11}$ reveals another step-like elastic anomaly for pressures exceeding 0.65 GPa. The anomaly and its dependence on frequency are discussed in detail below (Fig. 3 (b)).

\begin{figure*}
\includegraphics[width=0.8\linewidth]{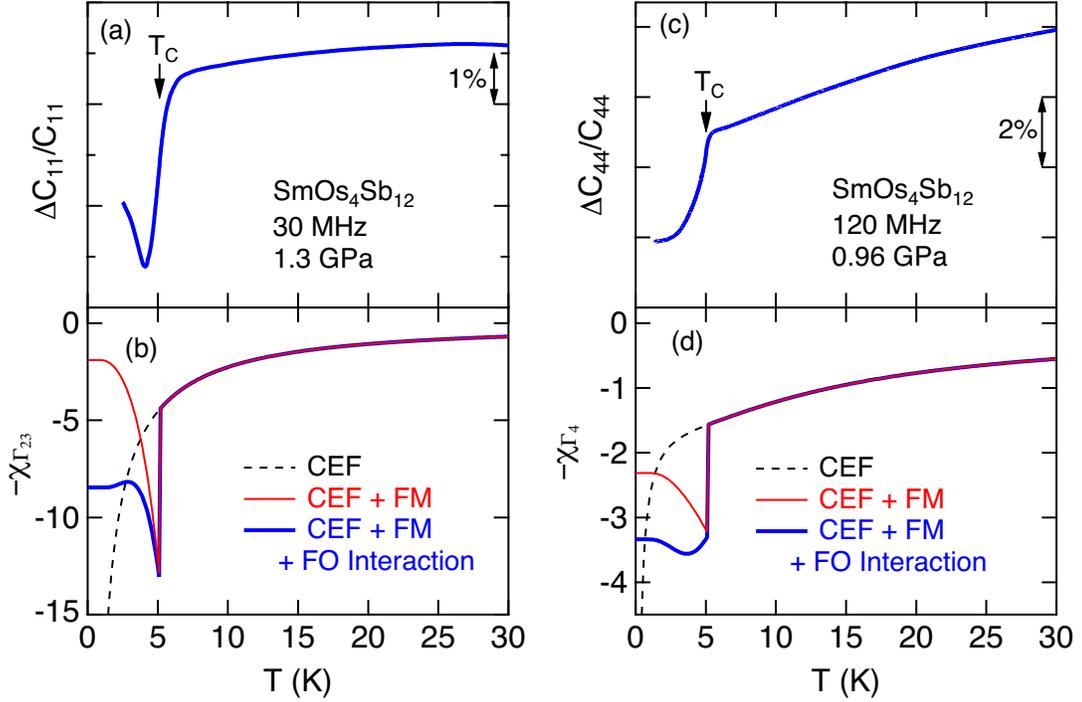}
\caption{\label{fig:epsart} (a) Temperature dependence of $C_{11}$, which corresponds to the response of the quadrupole $O_{20}$ with $\Gamma_{23}$ symmetry, at 1.3 GPa, (b) quadrupolar susceptibility $-\chi_{\Gamma_{23}}$ calculated by applying a mean-field approximation (see text),  (c) temperature dependence of $C_{44}$, which corresponds to the response of the quadrupole $O_{yz}$ with $\Gamma_{4}$ symmetry, at 0.96 GPa; and (d) quadrupolar susceptibility $-\chi_{\Gamma_{4}}$ calculated by applying the mean-field approximation.}
\end{figure*}

The insets of Fig. 2 show a phase diagram and the amount of the elastic anomalies ${\Delta}C$ of $C_{11}$ and $C_{44}$ at $T_{\rm C}$ as functions of pressure.
The anomalies at $T_{\rm C}$ of both elastic constants become larger at higher pressure, which is also explained by a recovery of the localized character of $4f$ electrons. This drastic change would have the same origin as the disappearance of the nuclear quadrupole resonance signal over a wide range at 0.4 GPa\cite{NQR_JPSJ_sa}.
Moreover, the peak in ac-susceptibility at $T_{\rm C}$ increases rapidly upon applying approximately 0.5 GPa \cite{UnderP_C_AcChi}. These results imply that the ordered state changes drastically from itinerant FM order to localized FM order upon applying a pressure greater than 0.4 GPa.
As shown by the phase diagram in the inset of Fig. 2, the pressure at which $T^{\ast}$ drops below $T_{\rm C}$ can be estimated by extrapolation to be 0.7 GPa. Note that this pressure is close to the pressure region where the exponent $n$ of the temperature dependence of electrical resistivity is a minimum \cite{NQR_PRL}.
Thus, these results strongly suggest that the NFL characteristic is related to the change in the localized character of the 4$f$ electrons.

We now discuss the ordered state in SmOs$_4$Sb$_{12}$ at high pressure, where the 4$f$ electrons are well localized, based on a comparison of the experimental results with a calculation based on the mean-field approximation. Because weak FM ordering appears at ambient pressure, we begin the calculation by considering a primary FM order. We also consider the intersite interaction of the $\Gamma_4$ magnetic-octupole moments $T^{\beta}_i = \overline{(\vector{J}_i\vector{J}_j^2-\vector{J}_k^2\vector{J}_i)}$, where $(i,j,k) = (x,y,z), (z,x,y)$, or $(y,z,x)$, and the bar represents normalized symmetrization. The Hamiltonian is
\begin{eqnarray}
H=H_{\rm CEF}-K_{\rm DD}{\langle}\vector{J}{\rangle}\vector{J}-K_{\rm OO}{\langle}T^{\beta}{\rangle}T^{\beta} - g_{\Gamma}O_{\Gamma}\varepsilon_{\Gamma},
\end{eqnarray}
where the first term corresponds to the CEF effect, and the coefficients $K_{\rm DD}$, $K_{\rm OO}$ and $g_{\Gamma}$ are the magnetic dipole-dipole coupling constant, magnetic octupole-octupole coupling constant and the strain-quadrupole coupling constant with symmetry ${\Gamma} = {\Gamma}_{23}$ or ${\Gamma}_4$, respectively. 
In this calculation, we assume $K_{\rm DD}$ = 2.2 and $K_{\rm OO}$ = 0.03.
Because the $\Gamma_{67}$ quartet ground state requires the magnetic easy axis for $\langle$100$\rangle$, we assume FM ordering along $\langle$100$\rangle$. In this calculation, we also consider three possible domain states by averaging over them. 
Figure 4 shows the experimental results obtained at the highest pressures and the calculated quadrupolar susceptibilities.
The quadrupolar susceptibility $-\chi_{{\Gamma}_{23}}$ calculated by assuming primary FM ordering agrees qualitatively with the experimentally obtained $C_{11}$. However, the same calculation for the $C_{44}$ mode gives a minimum $C_{44}$ at $T_{\rm C}$, and at low temperatures, $C_{44}$ increases with decreasing temperature, which is inconsistent with the experimental result. 
As shown in Fig. 4, if a ferro-type octupole intersite interaction is included in the calculation, the calculated quadrupolar susceptibility reproduces the leveling-off feature of $C_{44}$ at low temperatures.

Therefore, we conclude that the octupole-octupole interaction plays a role in the ordered state in this material. 
Considering the results of the specific heat and ac susceptibility under pressure \cite{UnderP_C_AcChi}, as discussed above, we strongly expect this interaction to become dominant and, at higher pressure, the ordered state to change into an octupolar ordered state . 

Next, we describe how the UDs depend on pressure. Figure 3(b) shows how $C_{11}$ and the ultrasonic attenuation ${\Delta}{\alpha}$ depend on temperature at 1.3 GPa and at various frequencies. As can be seen, a clear UD also appears under high pressures. The UD that appears at higher temperatures, which we call UD(high), shifts to lower temperatures upon increasing the pressure to 0.96 GPa. As seen in Ref. \onlinecite{SmOs4Sb12_Yana}, UD in $R$Os$_{4}$Sb$_{12}$ systems appears at higher temperatures as the atomic number increases.
In other words, as the ionic mass increases and the radius of the $R$ ion decreases, the temperature at which the UD appears shifts toward higher temperatures, whereas the size of the atomic cage does not change in the $R$Os$_{4}$Sb$_{12}$ systems. By applying pressure, the atomic cage gets smaller and the relative size of the guest ion increases. Thus, we expect that, under pressure, the UD shifts to lower temperatures, as observed below 0.96 GPa.
However, at 1.3 GPa, the elastic anomaly of UD(high) at 250 MHz is broader and appears at higher temperatures.
This result cannot be explained on the basis of the analogy above and suggests a possible relationship to the other parameters, such as the recovery of the localized character and the change of the mixed-valence state.
As previously noted, the temperature dependence of the averaged valence is reported to change upon applying pressure, and the temperature at which the averaged valence starts to decrease has a minimum around 1 GPa \cite{Valence_Pdep}.
The ion size also strongly depends on the valence. Thus, UD(high) is affected by such a change in the valence state.

The UD that appears at lower temperatures, which we call UD(low), gradually disappears with increasing pressure. Interestingly, this UD(low) seems to disappear at the pressure where $T^{\ast}$ drops below $T_{\rm C}$.
This result implies a possible relationship between the UD(low) and the $c$--$f$ hybridization. 
To verify the veracity of this relationship, further ultrasonic measurements of $C_{11}$ at high-pressure and high magnetic fields are needed, because a recovery of the localized character of 4$f$ electrons was also indicated by our ultrasonic measurements above 20 T at ambient pressure \cite{SmOs4Sb12_Pulse}.

Below $T_{\rm C}$, we report in the inset of Figs. 3(a) and 3(b), the two frequency-dependent step-like elastic anomalies at high pressures, which are accompanied by sharp peaks in ultrasonic absorption. In contrast, $C_{44}$ does not exhibit this type of anomaly in the region of pressure investigated in this work ($i$.$e$., these anomalies are mode selective, as are the UDs). Thus, these anomalies seem to have the same origin as the UDs.
Assuming the resonance relation ${\omega}{\tau} \sim 1$ is satisfied at the temperature where the absorption peaks, the relaxation time ${\tau}$ between these two anomalies seems to diverge below $T_{\rm C}$.
This result suggests that some phonon modes freeze out in this range of temperature and pressure, which couples to ${\Gamma}_1$ and/or ${\Gamma}_{23}$ strain waves.
We also observed a similar dynamical elastic response at 0.96 and 0.65 GPa.
Interestingly, this anomaly seems to appear under high pressure where the UD(low) disappears.
These experimental results for the frequency-dependent and mode-selective elastic anomalies lead us to the following simple conclusion: the anharmonic potential of the guest Sm ion, which may be causing the UD(low), might change as the character of the 4$f$ electrons changes from itinerant to localized upon applying pressure.
The origin and quadripartite relationship between $c$--$f$ hybridization, valence fluctuations, local charge fluctuations due to rattling, and multipolar interactions remain questions for discussion. Further measurements at higher pressure and over a wide range of temperature and frequency are required to elucidate these remaining issues.

In summary, we report high-frequency and high-pressure ultrasonic measurements of the filled skutterudite SmOs$_4$Sb$_{12}$. We detect a signature of the crossover from the itinerant 4$f$-electron region to localized 4$f$-electron region near 0.7 GPa. We also confirm the ${\Gamma}_{67}$ CEF ground state and the presence of octupolar degrees of freedom of the 4$f$ electrons in this material. We also measured how the UDs due to rattling depend on ultrasonic frequency and pressures.
The dependence of UDs on pressure implies that a relationship exists between the local charge fluctuation due to off-center degrees of freedom regarding rattling and the exotic 4$f$-electronic states of this compound. 

\begin{acknowledgments}
This research was supported by JSPS KAKENHI Grant No. 26400342, and 15K05882 and 15K21732(J-Physics), and the Strategic Young Researcher Overseas Visits Program for Accelerating Brain Circulation from the Japan Society for the Promotion of Science, and also supported by the U. S. DOE (Grant No. DEFG02-04-ER46105) and the U. S. NSF (Grant No. DMR-1506677).
We would like to thank Mr. Soichi Kasai (Coax. Co., Ltd.) for his technical assistance with the ultra-thin coaxial cable.
\end{acknowledgments}

\end{document}